\documentclass[showpacs,prl,10pt,twocolumn,superscriptaddress,aps]{revtex4-1}

\usepackage{ae}
\usepackage{bm} 
\usepackage{xcolor}
\usepackage{amsmath}
\usepackage{amssymb}
\usepackage{amsfonts}
\usepackage{graphicx}
\usepackage{slashed}
\usepackage{wasysym}
\usepackage{units}

\newcommand{\Eqref}[1]{Eq.~\eqref{#1}}
\newcommand{\visible}{discernible}
\newcommand{\visibility}{discernibility}
\newcommand{\vis}{dis}

\makeatletter
\def\fps@figure{ht}
\def\fps@table{ht}
\makeatother

\allowdisplaybreaks

\begin{document}

\setlength{\unitlength}{1mm}

\title{Boosting  quantum vacuum signatures by coherent harmonic focusing}

\author{Felix Karbstein}
\affiliation{Helmholtz-Institut Jena, Fr\"obelstieg 3, 07743 Jena, Germany}
\affiliation{Theoretisch-Physikalisches Institut, Abbe Center of Photonics, Friedrich-Schiller-Universit\"at Jena, Max-Wien-Platz 1, 07743 Jena, Germany}
\author{Alexander Blinne}
\affiliation{Helmholtz-Institut Jena, Fr\"obelstieg 3, 07743 Jena, Germany}
\author{Holger Gies}
\affiliation{Helmholtz-Institut Jena, Fr\"obelstieg 3, 07743 Jena, Germany}
\affiliation{Theoretisch-Physikalisches Institut, Abbe Center of Photonics, Friedrich-Schiller-Universit\"at Jena, Max-Wien-Platz 1, 07743 Jena, Germany}
\author{Matt Zepf}
\affiliation{Helmholtz-Institut Jena, Fr\"obelstieg 3, 07743 Jena, Germany}
\affiliation{Institut f\"ur Optik und Quantenelektronik, Abbe Center of Photonics, Friedrich-Schiller-Universit\"at Jena, Max-Wien-Platz 1, 07743 Jena, Germany}

\date{\today}

\begin{abstract}
 We show that coherent harmonic focusing provides an efficient mechanism to boost all-optical signatures of quantum vacuum nonlinearity in the collision of high-intensity laser fields, thereby offering a promising route to their first experimental detection.
 Assuming two laser pulses of given parameters at our disposal, we demonstrate a substantial increase of the number of signal photons measurable in experiments where one of the pulses undergoes coherent harmonic focusing before it collides with the fundamental-frequency pulse.
 Imposing a quantitative criterion to discern the signal photons from the background of the driving laser photons and accounting for the finite purity of polarization filtering,
 we find that signal photons arising from \textit{inelastic} scattering processes constitute a promising signature.
 By contrast, quasi-elastic contributions which are conventionally assumed to form the most prospective signal remain background dominated.
 Our findings may result in a paradigm shift concerning which photonic signatures of quantum vacuum nonlinearity are accessible in experiment.
\end{abstract}

\maketitle

\paragraph{Introduction}

The quantum vacuum has remarkable properties.
It is not trivial and inert, but amounts to a complex state whose properties are fully determined by quantum fluctuations. 
As these fluctuations comprise all existing particles, the quantum vacuum even constitutes a portal to new physics beyond the Standard Model of particle physics.
To obtain a measurable response, the quantum vacuum has to be probed by some external stimulus.
A powerful means is provided by strong macroscopic electromagnetic fields which couple directly to the charged particle sector.
Within the Standard Model, the leading effect arises from the effective coupling of the prescribed electric $\vec{E}$ and magnetic $\vec{B}$ fields via a virtual electron-positron pair. This process is governed by quantum electrodynamics (QED) and supplements Maxwell's classical theory in vacuum with effective nonlinear couplings of the electromagnetic fields \cite{Euler:1935zz,Heisenberg:1935qt,Weisskopf,Schwinger:1951nm}; see also \cite{Dittrich:2000zu,Dunne:2004nc,Marklund:2008gj,Dunne:2008kc,Heinzl:2008an,DiPiazza:2011tq,Dunne:2012vv,Battesti:2012hf,King:2015tba,Karbstein:2016hlj,Battesti:2018bgc}.

Up to now, the corresponding deviations have never been directly observed for macroscopically controlled fields.
This is because the effective self-interactions are parametrically suppressed by powers of $|\vec{E}|/E_{\rm cr}$ and $|\vec{B}|/B_{\rm cr}$, with critical electric (magnetic) field $E_\text{cr}=m_e^2c^3/(e\hbar)\simeq 1.3 \times 10^{18}\,\frac{\rm V}{\rm m}$ ($B_\text{cr}=E_\text{cr}/c \simeq 4 \times 10^{9}\,{\rm T}$).
The strongest macroscopic fields available in the laboratory are delivered by high-intensity lasers reaching peak fields $E\simeq{\cal O}(10^{14})\frac{\rm V}{\rm m}$ and $B\simeq{\cal O}(10^6){\rm T}$.
While these fields clearly fulfill $|\vec{E}|\ll E_{\rm cr}$, $|\vec{B}|\ll B_{\rm cr}$,
they appear to be sufficient to facilitate a first detection of QED vacuum signatures.
The basic idea is to collide high-intensity laser pulses and to look for vacuum-fluctuation-induced modifications of their properties, encoded in signal photons whose kinematics or polarization properties differ from the laser photons driving the effect, thereby allowing for a clear signal-to-background separation.
For recent estimates of the prospective numbers of signal photons attainable in laser pulse collisions, cf., e.g., Refs.~\cite{Lundstrom:2005za,Lundin:2006wu,Tommasini:2009nh,Tommasini:2010fb,King:2013am,King:2012aw,Gies:2017ygp,Gies:2017ezf,King:2018wtn,Blinne:2018nbd,Aboushelbaya:2019ncg}.
The smallness of the signal makes its detection challenging, even at dedicated high-intensity laser facilities such as CILEX \cite{CILEX}, CoReLS \cite{CoReLS}, ELI \cite{ELI} and SG-II \cite{SG-II}.

In this letter, we show that the number of attainable and, in particular, {\it \visible} signal photons can be increased significantly for a given laser pulse energy put into the interaction volume.
To this end, we rely on the mechanism of coherent harmonic focusing (CHF), pioneered by Refs.~\cite{Gordienko:2004,Gordienko:2005zz}.
Our quantitative analysis relies on the novel numerical approach \cite{Blinne:2018nbd} allowing for first-principles simulations of photonic signatures of vacuum nonlinearities.
We also provide analytical estimates based on a description of the driving laser fields as pulsed paraxial beams; cf. Ref.~\cite{Karbstein:2018omb}.

References~\cite{Gordienko:2004,Gordienko:2005zz} demonstrated that CHF can pave the way towards extreme intensities, thereby allowing for unprecedented experimental studies of nonperturbative electron-positron pair production in a highly localized strong field region; cf. also Ref.~\cite{Gonoskov:2013ada}.
They showed that the reflection of a relativistically intense laser pulse of wavelength $\lambda$ from the oscillating boundary of an overdense plasma produces a harmonic spectrum with the spectrum intensity scaling as $I_n\sim n^{-5/2}$, where $n\geq1$ labels the $n$th harmonic \cite{Gordienko:2004}.
These harmonics can be focused coherently down to a spot size of about $\lambda/n$ using a concave plasma surface of appropriate curvature \cite{Gordienko:2005zz}. 
While an improved description of the process resulted in a slight revision of the power as $5/2\to8/3$ \cite{Baeva:2006}, in this letter we stick to the original prediction of \cite{Gordienko:2004}.

As a concrete example, we employ CHF to boost photonic signatures of QED vacuum nonlinearity in the head-on collision of two linearly polarized high-intensity laser fields of given parameters.
For definiteness, we assume the initial laser pulses to agree in both wavelength $\lambda$ and pulse duration $\tau$.
One comprises an energy $W$ and is focused to a beam waist of $w_0=\lambda$. The other is reflected at a concave overdense plasma surface, effectively partitioning the laser pulse energy -- which after the reflection process is also assumed to be given by $W$ -- as $W=\sum_{n=1}^{n_\text{max}} W_n$ into the individual harmonics. Here, $W_n=Wn^{-5/2}/H^{(5/2)}_{n_{\rm max}}$ is the energy put into the $n$th harmonic, $n_{\rm max}$ is the harmonic cutoff and $H^{(q)}_{n_{\rm max}}=\sum_{n=1}^{n_\text{max}}1/n^{q}$ is a generalized harmonic number.
Experimentally, it has been shown that second harmonic generation achieves efficiencies of $22\pm8\%$ for intensities approaching $10^{21}\frac{\rm W}{{\rm cm}^2}$ [46].
These efficiencies are in good agreement with the scaling assumed for the CHF efficiency $2^{-5/2}=18\%$ for the n=2 channel and indicate the principle feasibility of implementing CHF scenarios with low harmonics in experiments.

The plasma surface focuses the $n$th harmonic to a waist of $w_{0,n}=\lambda/n$, such that the electric peak field amplitude of the $n$th harmonic scales as $E_{0,n}\sim \sqrt{W_n/(\tau w_{0,n}^2)}\sim n^{-1/4}$.
This CHF pulse collides head-on with the fundamental-frequency pulse at zero impact parameter and temporal offset in the focus. For the impact of spatio-temporal offsets on photonic quantum vacuum signatures, cf., e.g., Refs.~\cite{King:2012aw,Dinu:2014tsa,Karbstein:2016lby,Karbstein:2018omb}.

\paragraph{Formalism}

The amplitude for emission of a single signal photon (wave vector
$\vec{k}$, polarization $p$) from the electromagnetized QED vacuum reads \cite{Karbstein:2014fva}
\begin{equation}
 {\cal S}_{(p)}(\vec{k})\equiv\big\langle\gamma_p(\vec{k})\big| 
\Gamma_\text{int}[A(x),a(x)] \big|0\big\rangle \,. \label{eq:Sp}
\end{equation}
Here, $|\gamma_{p}(\vec{k})\rangle\equiv a^\dag_{\vec{k},p}|0\rangle$ is the 
single signal photon state and $\Gamma_\text{int}[A(x),a(x)]$ encodes the vacuum-fluctuation-mediated interactions of the operator-valued signal photon field $a(x)$ \cite{Karbstein:2014fva} with the driving macroscopic electromagnetic field $A(x)$ treated as a classical background \cite{Gies:2016yaa}.
For fields of frequencies $\omega\ll\frac{m_ec^2}{\hbar}$, these effective interactions are governed by the one-loop Heisenberg-Euler effective Lagrangian ${\cal L}_\text{HE}^{1\text{-loop}}$ \cite{Heisenberg:1935qt}, implying
\begin{equation}
 \Gamma_\text{int}[A(x),a(x)]\simeq\int{\rm d}^4x\,a^\mu (x)\,j_\mu(x)\,,
 \label{eq:Gamma}
\end{equation}
where $j_\mu(x)=2\,\partial^\alpha\frac{\partial{\cal L}_\text{HE}^{1\text{-loop}}}{\partial F^{\alpha\mu}}$ sources the signal photons.
The above validity criterion is met for present and near-future high-intensity lasers of optical to x-ray frequencies.

In the Heaviside-Lorentz System and units $c=\hbar=1$, the leading contribution of ${\cal L}_\text{HE}^{1\text{-loop}}$ reads \cite{Euler:1935zz,Heisenberg:1935qt}
\begin{equation}
 {\cal L}_\text{HE}^{1\text{-loop}}\simeq\frac{m_e^4}{8\pi^2}\frac{1}{45}\Bigl(\frac{e}{m_e^2}\Bigr)^4[(\vec{B}^2-\vec{E}^2)^2+7(\vec{B}\cdot\vec{E})^2\bigr]\,.
 \label{eq:HEpert}
\end{equation}
Equation~\eqref{eq:HEpert} is valid for $|\vec{E}|\ll E_{\rm cr}$, $|\vec{B}|\ll B_{\rm cr}$ and should allow for the reliable study of all-optical signatures of QED vacuum nonlinearity driven by high-intensity lasers with an accuracy on the $1\%$ level \cite{Blinne:2018nbd}.

Upon insertion of \Eqref{eq:Gamma} into \Eqref{eq:Sp}, the signal photon emission amplitude can be expressed as
\begin{equation}
 S_{(p)}(\vec{k})=\frac{\epsilon_{(p)}^{*\mu}(\vec{k})}{\sqrt{2k^0}}\int{\rm d}^4x\,{\rm e}^{{\rm i}kx}\,j_{\mu}(x)\,\biggr|_{k^0=|\vec{k}|}\,, \label{eq:Sp2}
\end{equation}
where $\epsilon_{(p)}^{\mu}(\vec{k})=(0,\vec{e}_{(p)}(\vec{k}))$, fulfilling $|\vec{e}_{(p)}(\vec{k})|=1$ and $\vec{k}\cdot\vec{e}_{(p)}(\vec{k})=0$, is the polarization vector of the signal photon state $|\gamma_{p}(\vec{k})\rangle$.
Here, we label the two polarizations transverse to $\vec{k}$ by $p\in\{1,2\}$.
Without loss of generality, we choose the polarization basis such that the
polarization vector $\vec{e}_{(1)}(\vec{k})$ always fulfills
$\vec{e}_{(1)}(\vec{k})\cdot\vec{\epsilon}_0=0$, i.e., is perpendicular to both $\vec{k}$ and a given constant reference vector $\vec{\epsilon}_0$. Hence, $\vec{e}_{(1)}(\vec{k})$ spans the polarization mode polarized perpendicularly to $\vec{\epsilon}_0$, and $\vec{e}_{(2)}(\vec{k})$ is the vector completing the orthogonal basis.

With these definitions, \Eqref{eq:Sp2} yields
\begin{align}
 {\cal S}_{(p)}(\vec{k})={\rm i}\sqrt{\frac{\rm k}{2}}\int&{\rm d}^4x\, {\rm e}^{{\rm i}(\vec{k}\cdot\vec{x}-{\rm k}t)} \nonumber\\
 \times\bigl[&\vec{e}_{(p)}(\vec{k})\cdot\vec{P}-\vec{e}_{(p+1)}(\vec{k})\cdot\vec{M}\bigr],
 \label{eq:Sp1}
\end{align}
with ${\rm k}=|\vec{k}|$ and $\vec{e}_{(3)}(\vec{k})=-\vec{e}_{(1)}(\vec{k})$. The polarization $\vec{P}$ and magnetization $\vec{M}$ of the quantum vacuum are defined as \cite{Berestetskii}
\begin{equation}
 \vec{P}=\frac{\partial{\cal L}_{\rm HE}^{1\text{-loop}}}{\partial\vec{E}}\quad\text{and}\quad\vec{M}=-\frac{\partial{\cal L}_{\rm HE}^{1\text{-loop}}}{\partial\vec{B}}\,.
\end{equation}
Finally, the differential number of signal photons of polarization $p$ is related to the modulus square of \Eqref{eq:Sp1} and reads
\begin{equation}
{\rm d}^3N_{(p)}(\vec{k})=\,\frac{{\rm d}^3k}{(2\pi)^3}\bigl|{\cal
S}_{(p)}(\vec{k})\bigr|^2 \,. \label{eq:d3Np_polarcoords}
\end{equation}
For a polarization insensitive measurement we have ${\rm d}^3N=\sum_{p=1}^2{\rm d}^3N_{(p)}$.

\paragraph{Field configuration}

To describe the electromagnetic fields of a focused laser pulse (wavelength $\lambda$, energy $W$, duration $\tau$, waist $w_0=\lambda$), we employ the spectral pulse model \cite{Waters:2017tgl}, detailed in Sec.~III~D~2 of Ref.~\cite{Blinne:2018nbd}.
These fields fulfill Maxwell's equations in vacuum exactly and are conveniently represented in terms of a complex vector potential in radiation gauge, 
\begin{align}
 \vec{A} (x) = \int\mathop{\frac{{\rm d}^3{k}}{\left( 2\pi \right)^3}} {\rm e}^{{\rm i}(\vec{k} \cdot \vec{x}-{\rm k}t)}\,\sum_{q=1}^2 \vec{e}_{(q)} (\vec{k})\,a_q(\vec{k}) \,,
 \label{eq:A}
\end{align}
with spectral amplitudes $a_q(\vec{k})$ encoding the spatio-temporal field structure.
The associated real-valued electric (magnetic) field is given by $\vec{E}(x)=\Re\{-\partial_t\vec{A}(x)\}$  ($\vec{B}(x)=\Re\{\vec{\nabla}\times\vec{A}(x)\}$).
For a laser pulse propagating in $\pm\hat{\vec{\kappa}}$ direction, polarized along $\vec{\epsilon}_\pm$ in the focus at $x^\mu=0$, the spectral amplitudes are given by $a_q(\vec{k})\to a_q^\pm(\vec{k})$,
\begin{align}
  a_q^\pm(\vec{k}) =&\pm\frac{(2\pi)^{\frac{3}{4}}}{{\rm i}{\rm k}}\,\vec{\epsilon}_\pm\cdot\vec{e}_{(q)}(\vec{k}) \,\Theta(\pm k_\parallel) \frac{k_\parallel}{\rm k}
\nonumber\\
  &\times \sqrt{W\tau}\,\lambda\,{\rm e}^{-(\frac{\lambda}{2})^2{\rm k}_\perp^2 - (\frac{\tau}{4})^2[{\rm k}-\omega(\lambda)]^2} \,. \label{eq:aq}
\end{align}
Here, $\Theta(.)$ denotes the Heaviside function, $\omega(\lambda)=\frac{2\pi}{\lambda}$ the laser photon energy, $k_\parallel=\hat{\vec{\kappa}}\cdot\vec{k}$ the momentum component along $\hat{\vec{\kappa}}$, and ${\rm k}_\perp=\sqrt{{\rm k}^2-k_\parallel^2}\geq0$.
The amplitudes~\eqref{eq:aq} have been constructed such that the zeroth-order paraxial Gaussian beam is reproduced for weak focusing and long pulse durations \cite{Waters:2017tgl,Karbstein:2017jgh}.

To model our scenario of a fundamental-frequency laser pulse (propagation direction $\hat{\vec{\kappa}}$, polarization $\vec{\epsilon}_+$ in the focus) colliding head-on with a CHF pulse (polarization $\vec{\epsilon}_-$ in the focus) containing $n_\text{max}$ harmonics, we choose the spectral amplitudes in \Eqref{eq:A} as
\begin{equation}
 a_q(\vec{k}) \to a_q^+(\vec{k}) +\sum_{n=1}^{n_\text{max}}\sqrt{\frac{W_n}{W}}\, a_q^-(\vec{k})\big|_{\lambda\to\frac{\lambda}{n}}\,.
 \label{eq:spec_am}
\end{equation}
Subsequently, we refer to the laser pulse propagating in $\pm\hat{\vec{\kappa}}$ direction as ``$\pm$'' pulse; the `$`+$'' (``$-$'') pulse is the fundamental-frequency (CHF) pulse.

\begin{figure}
 \centering
 \includegraphics[width=\linewidth]{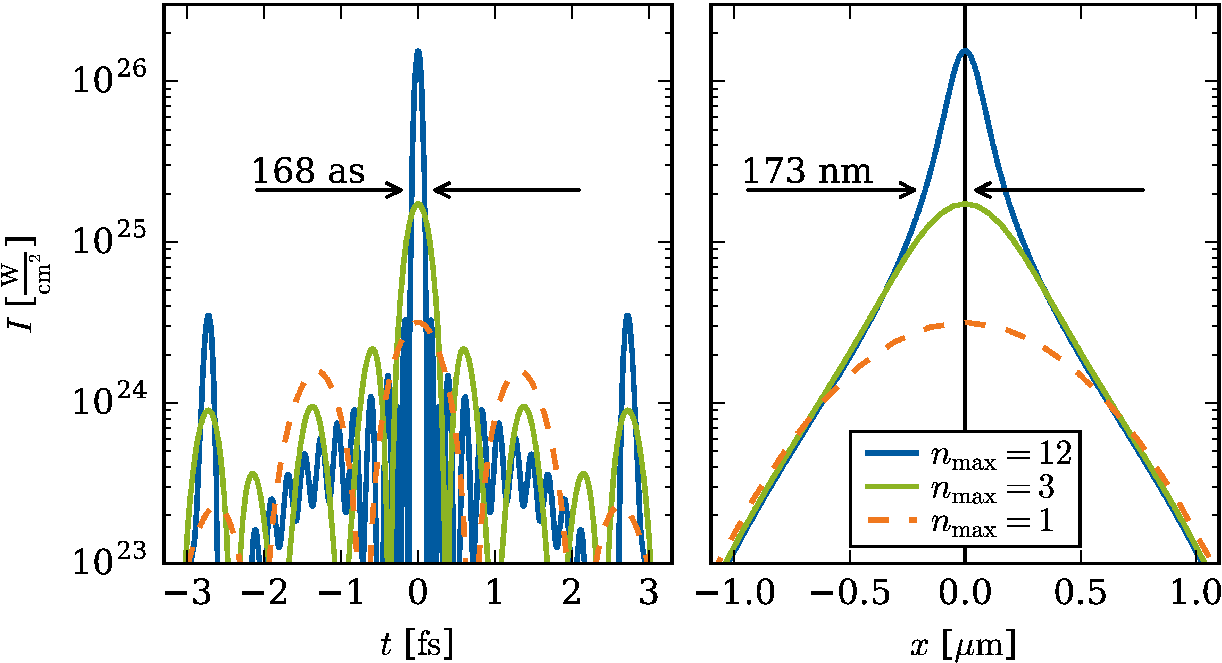}
 \caption{Characteristics of the CHF pulse with total energy $W=25\,{\rm J}$, envelope $\tau=5\,{\rm fs}$ and fundamental wavelength $\lambda=800\,{\rm nm}$ for different $n_{\rm max}$.
 Left: temporal profile in the focus. Right: transverse focus profile.
 The energy $W$ is partitioned into $n_{\rm max}$ harmonics of wavelength $\lambda_n=\lambda/n$ and energy $W_n$.
 Each mode is focused to its diffraction limit $w_{0,n}=\lambda_n$.
 For $n_{\rm max}=12$, the effective $1/{\rm e}^2$ pulse duration and waist of the CHF pulse are as small as $\tau_{\rm CHF}=168\,{\rm as}$ and $w_{\rm CHF}=173\,{\rm nm}$.}
 \label{fig:chfProfiles}
\end{figure}
The coherent superposition of the $n_{\rm max}$ diffraction limited harmonics to form the CHF pulse results in a narrow, strongly peaked pulse shape of effective waist $w_{\rm CHF}\approx \lambda/n_{\rm max}$ \cite{Gordienko:2005zz}.
For large values of $n_{\rm max}$, the effective pulse duration in the focus $\tau_{\rm CHF}$ becomes essentially independent of the envelope $\tau$ of the contributing modes.
Instead, it is also determined by the wavelength and given by $\tau_{\rm CHF}\approx \lambda/n_{\rm max}$ \cite{Gordienko:2005zz}; cf. Fig.~\ref{fig:chfProfiles}.

\paragraph{Results}

In the remainder, we use the following parameters: $\lambda=800\,{\rm nm}$, $\tau=5\,{\rm fs}$ and $W=25\,{\rm J}$.
Both pulses are linearly polarized; the angle between their polarization vectors in the focus is $\phi=\sphericalangle(\vec{\epsilon}_+,\vec{\epsilon}_-)$.
The value of $\tau=5\,{\rm fs}$ is chosen mainly for numerical convenience, allowing us to scale $n_{\rm max}$ up to $12$.
Such small pulse durations were so far only achieved at sub-Joule pulse energies \cite{Laszlo:2017}; state-of-the-art high-intensity laser pulses feature durations $\gtrsim20\,{\rm fs}$ \cite{ELI}.
We have explicitly confirmed for $n_{\rm max}=6$ and ELI-NP \cite{ELI} parameters ($\lambda=800\,{\rm nm}$, $\tau=20\,{\rm fs}$, $W=200\,{\rm J}$) that the studied effects persist for longer pulse durations; cf. the Supplementary Material.

As will be demonstrated below, to a very good approximation the signal photons $N^\pm_{(p)}$ emitted into the ``$\pm$'' half-space, characterized by wave vectors $\vec{k}$ fulfilling $\pm\hat{\vec{\kappa}}\cdot\vec{k}>0$,
can be interpreted as arising from the ``$\pm$'' pulse and being quasi-elastically scattered off the ``$\mp$'' pulse.
Manifestly inelastic scattering processes characterized by an energy transfer of ${\cal O}(\omega)$ are suppressed in comparison to the elastic contributions \cite{Karbstein:2015xra,Karbstein:2014fva,Gies:2017ygp}.
The study of photon scattering in the head-on collision of two linearly polarized paraxial beams \cite{Karbstein:2019bhp} suggests that an angle of $|\phi|=\frac{\pi}{2}$ between the polarization vectors $\vec{\epsilon}_\pm$ maximizes the signal photon number $N$ attainable in a polarization insensitive measurement.
By contrast, the number $N_\perp$ of signal photons scattered into a perpendicularly polarized ($\perp$) mode should become maximum for an angle of $|\phi|=\frac{\pi}{4}\,{\rm mod}\,\pi$.
We have explicitly confirmed this behavior in our simulations (see Fig.~1 in the Supplementary Material) and stick to these optimal choices of $\phi$ when providing results for $N$ and $N_\perp$ in the remainder.

Aiming at the analysis of the polarization-flipped signal photons propagating into the ``$\pm$'' half-space, we choose $\vec{\epsilon}_0\to\vec{\epsilon}_\pm$. This immediately implies that $\vec{e}_{(1)}(\vec{k})$ spans the mode polarized perpendicularly to $\vec{\epsilon}_\pm$, respectively, and $N_\perp:=N_{(1)}$. At the same time, this choice ensures that none of the driving laser photons $\cal N$ are perpendicularly polarized; cf. \Eqref{eq:aq}.

\begin{figure}
 \centering
 \includegraphics[width=\linewidth]{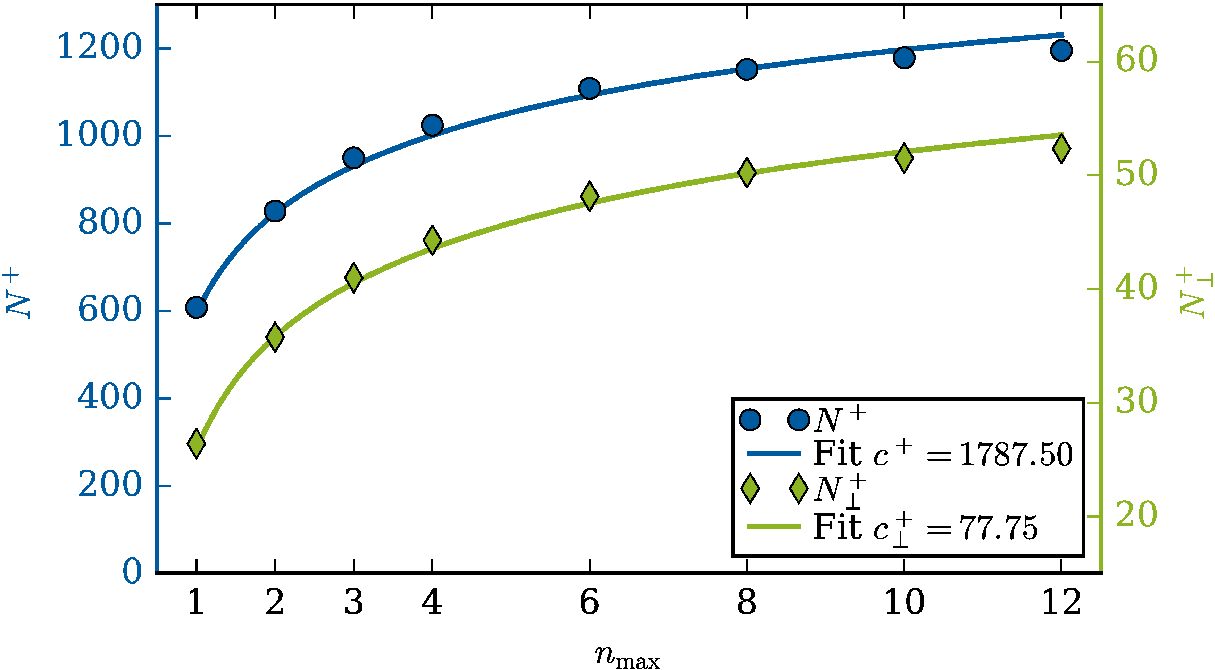}
 \caption{Scaling of the signal photon numbers $N^+$ (blue dots; left scale) and $N^+_\perp$ (green diamonds; right scale) with $n_{\rm max}$.
 The solid lines are least-squares fits to \Eqref{eq:scaling}.}
 \label{fig:nmaxNNperp}
\end{figure}

Figure~\ref{fig:nmaxNNperp} depicts the attainable numbers of signal photons $N^+$ and $N_\perp^+$ as a function of $n_{\rm max}$.
The simulation results are remarkably well described by 
\begin{align}
N^+_{(p)}(n_{\rm max})
 =c^+_{(p)}\frac{1}{n_{\rm max}(1+2n_{\rm max}^2)}\frac{(H^{(1/4)}_{n_{\rm max}})^4}{(H^{(5/2)}_{n_{\rm max}})^2}\,, \label{eq:scaling}
\end{align}
with polarization dependent numerical constants $c^+_{(p)}$. 
Equation~\eqref{eq:scaling} follows from Eqs.~(7) and (10) of Ref.~\cite{Karbstein:2018omb} upon identification of the probe (pump) with the fundamental frequency (CHF) pulse and assuming $w_{\rm CHF}=w_{0}/n_{\rm max}$ as well as $\tau_{\rm CHF}\sim{\rm z}_{R,{\rm CHF}}\sim 1/n_{\rm max}$. It only accounts for quasi-elastically scattered signal photons.
The CHF peak field energy per spot size is determined as $\sqrt{W_{\rm CHF}/w_{\rm CHF}^2}\to\sum_{n=1}^{n_{\rm max}}\sqrt{W_n/w_{0,n}^2}$.
Analytical estimates for the angular decay of $N^\pm_{(p)}$ as well as the radial divergences $\theta^\pm_{\rm sig}$ of the signal photons emitted into the ``$\pm$'' space can be derived along the same lines; see the Supplementary Material.

Equation~\eqref{eq:scaling} implies that CHF can increase $N_{(p)}^+$ at most by a factor of
\begin{equation}
 \frac{N^+_{(p)}(n_{\rm max}\gg1)}{N^+_{(p)}(n_{\rm max}=1)}\simeq\frac{128}{27\zeta^2(\frac{5}{2})}\approx2.6
 \label{eq:totalincrease}
\end{equation}
relative to the collision of two fundamental-frequency pulses.
Asymptotically, the increase of the CHF peak field with $n_{\rm max}$ is compensated by a decrease of the effective focusing volume.
The ratio of the coefficients $c^+$ and $c_\perp^+$ extracted in Fig.~\ref{fig:nmaxNNperp} is $c^+/c_\perp^+\approx23.0$, and thus roughly agrees with that found for counter-propagating paraxial beams $c^+/c_\perp^+=\frac{197}{9}\approx21.8$ \cite{Karbstein:2019bhp}.

\begin{figure}
 \centering
 \includegraphics[width=\linewidth]{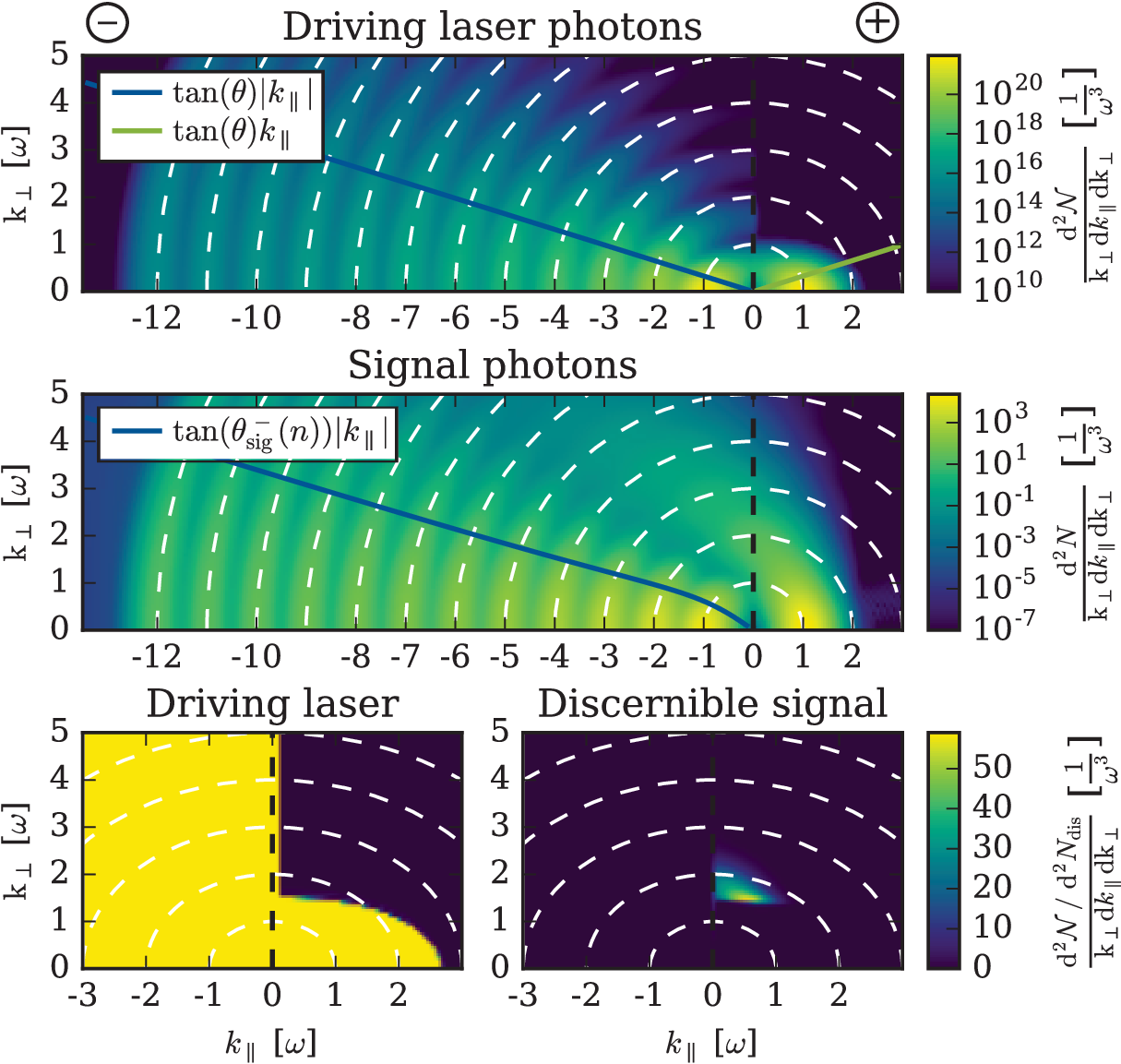}
 \caption{Spectra of the driving laser photons  $\cal N$ and signal photons $N$ attainable in a polarization insensitive measurement for $n_{\rm max}=12$. White dashed circles indicate lines of constant photon energy ${\rm k}=n \omega$, $n\in\mathbb{N}$. Different color scales are used in the top, middle and bottom panels.
 In the top panel $\theta=1/\pi$ denotes the radial divergence of a diffraction limited Gaussian beam.
 The radial divergence highlighted in the middle panel is determined from Eq.~(4) in the Supplementary Material.
 The bottom panels focus on the spectral domain where the differential number of signal photons surpasses the differential number of driving laser photons.
 Here, we compare the spectrum of the driving laser photons with the filtered signal photon spectrum fulfilling the criterion ${\rm d}^3N/{\rm d}^3k>{\rm d}^3{\cal N}/{\rm d}^3k$
 adopting the same linear color scale.
 Integrating the latter, we obtain $N_{\rm dis}\approx26.06$ \visible\ signal photons per shot.}
 \label{fig:specs}
\end{figure}

Subsequently, we focus on simulation data for the signal photon spectra.
In Fig.~\ref{fig:specs} we highlight the CHF case with $n_{\rm max}=12$.
For a simple and fair assessment of the benefits of CHF, we compare the results of this CHF scenario with those for the collision of two fundamental frequency pulses of the same energy ($n_{\rm max}=1$).
To assess the separability of the signal photons from the background, we also analyze the spectrum of the driving laser photons.
The integrated numbers of laser photons inferred from our simulation are
\begin{center}
 \begin{tabular}{c||c|c|}
 $n_{\rm max}$ & $1$ & $12$  \\
 \hline
 \hline
 ${\cal N}$  &  $2.0\times10^{20}$ &  $1.86\times10^{20}$ \\
 \hline
\end{tabular}
\end{center}
\noindent constituting the background from which the signal has to be separated. These results are in good agreement with the analytical estimates obtained with Eq.~(8) in the Supplementary Material.

We call the differential number of signal photons {\it \visible} from the background if
it fulfills the \textit{\visibility\ criterion} ${\rm d}^3N_{(p)}/{\rm d}^3k>{\rm d}^3{\cal N}_{(p)}/{\rm d}^3k$.
Summing over the two transverse polarizations $p$ and integrating over the spectral regions where this criterion holds, 
we obtain
\begin{center}
 \begin{tabular}{c||c|c|}
 $n_{\rm max}$ & $1$ & $12$  \\
 \hline
 \hline
 $N_{\rm \vis}$  & $2.15\times10^{-6}$  &  $26.06$ \\
 \hline
\end{tabular}
\end{center}
\noindent \visible\ signal photons per shot at $\approx2\omega$.
This implies that essentially none of the quasi-elastically scattered signal photons, which dominate the total numbers of signal photons $N_{(p)}$ (cf. \Eqref{eq:totalincrease} and Fig.~\ref{fig:nmaxNNperp}), can be discerned from the background of the driving laser photons.
Microscopically, the discernible signal photons at $\approx2\omega$ appear to arise from the merging of two counter-propagating fundamental-frequency laser photons in the localized strong field of the CHF pulse.
For an analogous scenario with ELI-NP \cite{ELI} laser parameters and $n_{\rm max}=6$, we obtain $N_{\rm dis}\approx314$ at $\approx2\omega$; see Fig.~3 in the Supplementary Material.

Specializing the above criterion to the $\perp$ polarization mode, we obtain
\begin{center}
 \begin{tabular}{c||c|c|}
 $n_{\rm max}$ & $1$ & $12$  \\
 \hline
 \hline
 $N_{\perp,{\rm \vis}}$ & $57.93$ & $151.31$ \\
 \hline
\end{tabular}\quad for\quad ${\cal P}=0$,
\end{center}
\noindent amounting to the polarization purity of an ideal polarization filter. These signal photons are predominantly emitted in ``$+$'' forward direction at $\approx\omega$.
This can be explained by the fact that the driving laser photons have zero overlap with the $\perp$ mode, allowing for an essentially background free measurement of the quasi-elastic scattering signal.
The gain of $\approx2.6$ achieved by CHF with $n_{\rm max}=12$ relative to $n_{\rm max}=1$ observed here is fully compatible with the gain of $\approx2.4$ derived for the quasi-elastically scattered signal photons in Eq.~(6) of the Supplementary Material.
The signal photons constituting $N_{\perp,{\rm \vis}}$ do in general not form a subset of $N_{{\rm \vis}}$, because both sides of the discernibility criterion are inherently polarization sensitive.
For a realistic polarization filter with ${\cal P}\neq0$ the discernibility criterion reads ${\rm d}^3N_\perp/{\rm d}^3k>{\cal P}\,{\rm d}^3{\cal N}/{\rm d}^3k$.
In this case, the gain achieved by CHF is substantially larger and the number of \visible\ signal photons using an ambitious polarization purity becomes
\begin{center}
 \begin{tabular}{c||c|c|}
 $n_{\rm max}$ & $1$ & $12$  \\
 \hline
 \hline
 $N_{\perp,{\rm \vis}}$ & $6.59\times10^{-3}$ & $10.44$ \\
 \hline
 \end{tabular}\quad for\quad ${\cal P}=10^{-10}$.
\end{center}
These signal photons again feature an energy of $\approx2\omega$.
The corresponding spectra qualitatively agree with those depicted in Fig.~\ref{fig:specs} upon identifying ${\cal N}\to{\cal P}\,{\cal N}$ and $N_{{\rm \vis}}\to N_{\perp,{\rm \vis}}$.

The results for $N_{\rm \vis}$ and $N_{\perp,{\rm \vis}}$ for ${\cal P}\neq0$, which are -- at least in principle -- accessible in experiment, clearly underpin the substantial enhancement of several orders of magnitude in the numbers of \visible\ signal photons achieved by CHF.
It is interesting to note that such a dramatic enhancement could not be expected from the comparably moderate increase of the total numbers of signal photons $N_{(p)}$ (cf. \Eqref{eq:totalincrease} and Fig.~\ref{fig:nmaxNNperp}) as well as the results for $N_{\perp,{\rm \vis}}$ based upon the existence of a perfect polarization filter with ${\cal P}=0$.
In this sense, our results exemplify that the criterion of the principle possibility of an experimental measurement of the effect based on real-world limitations, like a non-zero polarization purity, may significantly impact the assessment if the implementation of an advanced scheme, such as CHF, in experiment is worthwhile or not.

Our results exemplify that the assessment of an advanced scheme, such as CHF,
necessarily requires the consideration of real-world limitations. In the
present case, the mere existence of a rather small polarization purity leads
to a decisive change of perspective on the relevance of the CHF scheme. 

\paragraph{Conclusions}

We have demonstrated in an idealized setup that CHF can substantially increase the number of \visible\ signal photons in the collision of high-intensity laser pulses
for a given energy put into the interaction volume.
We are confident that our findings will stimulate many further theoretical ideas and proposals as well as dedicated experimental campaigns aiming at the first verification of quantum vacuum nonlinearity using CHF and replications based on conventional higher-harmonic generation techniques.

\acknowledgments

This work has been funded by the Deutsche Forschungsgemeinschaft (DFG) under Grant Nos. 416611371; 416607684; 416702141; 416708866 within the Research Unit FOR2783/1.

\onecolumngrid

\newpage\pagebreak

\setcounter{equation}{0}

\section*{Supplementary Material}

\begin{figure}
 \includegraphics[width=1\linewidth]{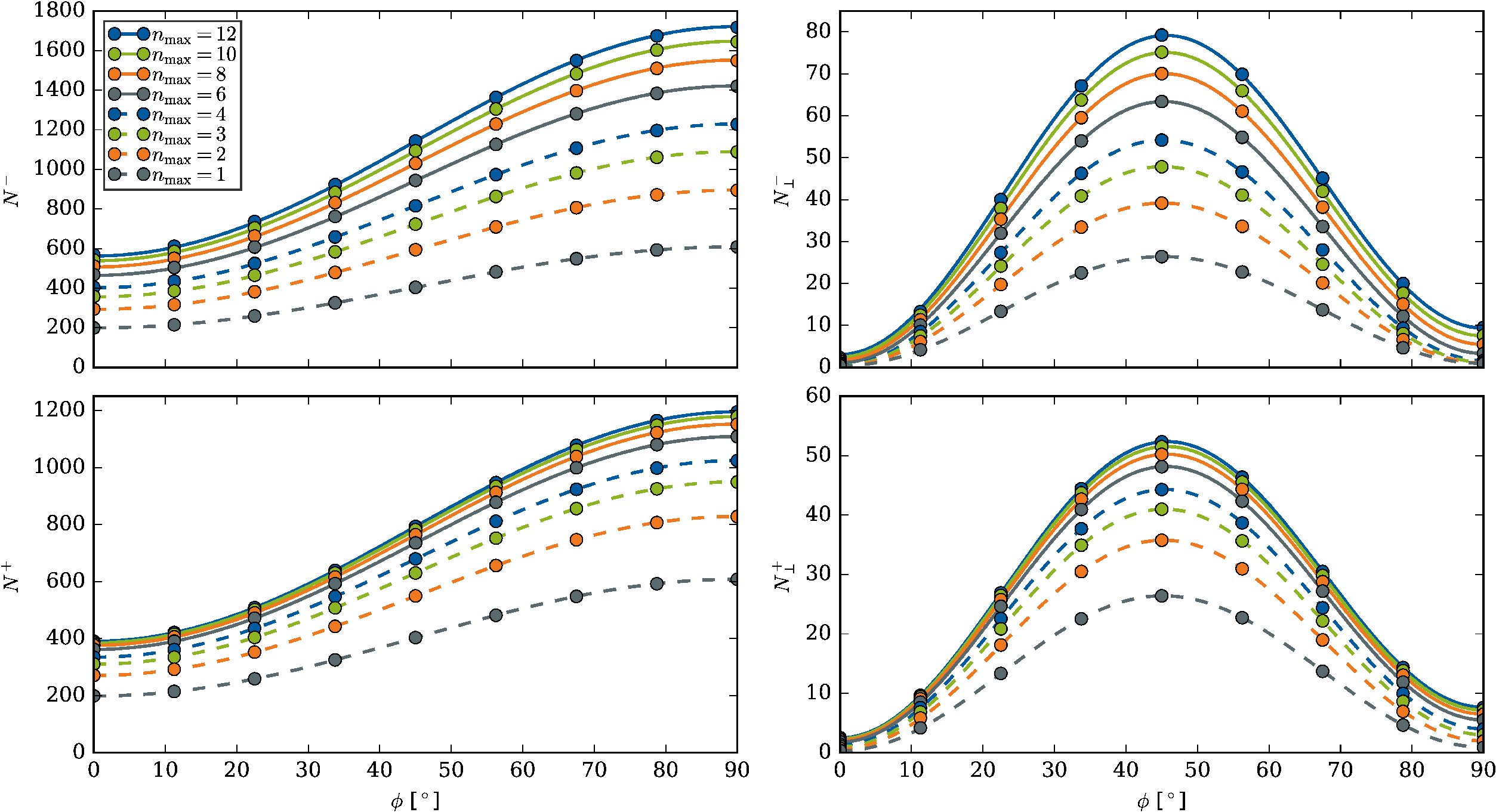}
 \caption{Numbers of signal photons $N$ attainable in a polarization insensitive measurement (left) and numbers of signal photons $N_\perp$ scattered into a perpendicularly polarized mode (right) for different values of $n_{\rm max}$ as a function of the relative angle between the polarization vectors of the driving laser pulses in the focus $\phi=\sphericalangle(\vec{\epsilon}_+,\vec{\epsilon}_-)$.
 The panels in the upper (lower) line show results for signal photons emitted into the ``$-$'' (``$+$'') direction. The number of attainable signal photons $N$ ($N_\perp$) is at a maximum for an relative angle of $\phi=90^\circ$ ($\phi=45^\circ$) between the polarization vectors $\vec{\epsilon}_\pm$.
 The lines are least squares fits of the functions $N^\pm(\phi) = A(130-66\cos(2\phi))$ and $N_\perp^\pm(\phi) =B(130-66\cos(2\phi)) + C\sin^2(2\phi)$, with fitting coefficients $A$, $B$ and $C$, to the simulation data points (filled circles).
 The first fitting function is modeled after the analytical result for $N(\phi)$ in the head-on collision of two paraxial beams \cite{Karbstein:2019bhp}.
 The latter amounts to a combination of the former and the corresponding paraxial result for $N_\perp(\phi)$ \cite{Karbstein:2019bhp}. For non-paraxial beams all signal photons $N$ generically exhibit a non-vanishing overlap with the $\perp$ mode \cite{Blinne:2018nbd} motivating this combination. This contribution is also essential in accounting for the asymmetry of the simulation data for $N^\pm_\perp(\phi)$, namely $N^\pm_\perp(\phi)\neq N^\pm_\perp(90^\circ-\phi)$.}
 \label{fig:Ns(beta)}
\end{figure}

Here we provide some additional material underpinning the arguments given in the main text.
Figure~\ref{fig:Ns(beta)} confirms that the total number of signal photons attainable in a polarization insensitive measurement $N$ and the number of signal photons scattered into a perpendicularly polarized mode $N_\perp$ reach their maxima for different choices of the relative angle between the polarization vectors of the driving laser pulses in the focus $\phi=\sphericalangle(\vec{\epsilon}_+,\vec{\epsilon}_-)$.
As detailed in the main text, considerations based on an analysis of the head-on collision of two paraxial laser fields predict $N$ ($N_\perp$) to be at a maximum for $|\phi|=\frac{\pi}{2}$ ($|\phi|=\frac{\pi}{4}\,{\rm mod}\,\pi$).
In the angle interval $\phi\in[0^\circ\ldots90^\circ]$ considered in Fig.~\ref{fig:Ns(beta)}, the corresponding angle is $\phi=90^\circ$ ($\phi=45^\circ$) for $N$ ($N_\perp$).
Our simulation data presented in Fig.~\ref{fig:Ns(beta)} are perfectly compatible with these predictions.

The accurate description of $N^+_{(p)}$ by Eq.~(12) in the main body of this letter suggests that Eq.~(4) of Ref.~\cite{Karbstein:2018omb} can serve as an analytical estimate for the angular decay of $N^+_{(p)}$. 
Sticking to the same assumptions, we obtain
\begin{align}
 \frac{{\rm d}N^+_{(p)}(n_{\rm max})}{\vartheta^+{\rm d}\vartheta^+}
 \simeq \frac{4}{\theta^2}\frac{c^+_{(p)}}{n_{\rm max}(1+2n_{\rm max}^{2})^2}\frac{(H^{(1/4)}_{n_{\rm max}})^4}{(H^{(5/2)}_{n_{\rm max}})^2}\,
  {\rm e}^{-2(\frac{\vartheta^+}{\theta})^2\frac{1}{1+2n_{\rm max}^{2}}}\,, \label{eq:N+sig}
\end{align}
where $\vartheta^+$ is the polar angle measured relative to the forward beam axis of the ``$+$'' pulse.
Equation~\eqref{eq:N+sig} fully characterized by $c^+_{(p)}$, $n_{\rm max}$ and the radial divergence of the driving laser pulses $\theta$.
For diffraction limited driving beams as considered here, we have $\theta=\frac{1}{\pi}$.
The corresponding estimate for the radial divergence of the signal photons emitted into the ``$+$'' half-space is
\begin{equation}
 \theta^+_{\rm sig}(n_{\rm max})\simeq\theta\,\sqrt{1+2n_{\rm max}^2}\,\xrightarrow{n_{\rm max}\gg1}\,\sqrt{2}\,n_{\rm max}\theta\,. \label{eq:theta+}
\end{equation}
Note that $\theta^+_{\rm sig}(n_{\rm max})\geq\sqrt{3}\theta$ generically surpasses the radial divergence $\theta$ of the ``$+$'' laser pulse.
Particularly for large values of $n_{\rm max}$ the result of \Eqref{eq:N+sig}, and thus also \Eqref{eq:theta+}, violate the condition of $\vartheta\ll1$ entering their derivation in Ref.~\cite{Karbstein:2018omb}.
Nevertheless, \Eqref{eq:theta+} indicates that the signal photons originating from the ``$+$'' pulse being quasi-elastically scattered off the strongly peaked ``$-$'' pulse are distributed in a much wider angular regime than the divergence $\theta$ of the driving laser pulses.

Simultaneously, photons of each of the $n$ modes constituting the ``$-$'' pulse experience quasi-elastic scattering off the ``$+$'' pulse. The analytical estimate for their angular decay is
\begin{align}
 \frac{{\rm d}N^-_{(p)}(n_{\rm max},n)}{\vartheta^-{\rm d}\vartheta^-}
 \simeq \frac{4}{\theta^2}\frac{c^+_{(p)}}{(2+n^2)^2}\frac{n^{5/2}}{H^{(5/2)}_{n_{\rm max}}}\, {\rm e}^{-2(\frac{\vartheta^-}{\theta})^2\frac{n^2}{2+n^2}}\,, \label{eq:N-sig}
\end{align}
where $\vartheta^-$ is the polar angle measured relative to the forward beam axis of the ``$-$'' pulse.
This expression also follows from Eq.~(4) of Ref.~\cite{Karbstein:2018omb}, taking into account the invariance
of our collision scenario under relabeling $\pm\to\mp$ for $n_{\rm max}=1$.
From~\Eqref{eq:N-sig} we infer
\begin{equation}
 \theta^-_{\rm sig}(n)\simeq\theta\,\sqrt{\frac{2+n^2}{n^2}}\,\xrightarrow{n\gg1}\,\theta\,, \label{eq:theta-}
\end{equation}
such that the signal photon contributions of higher modes are predominantly scattered into the forward cone of the ``$-$'' pulse. This result is in line with the assumption of $\vartheta\ll1$ entering its derivation \cite{Karbstein:2018omb}.

\begin{figure}
 \includegraphics[width=0.5\linewidth]{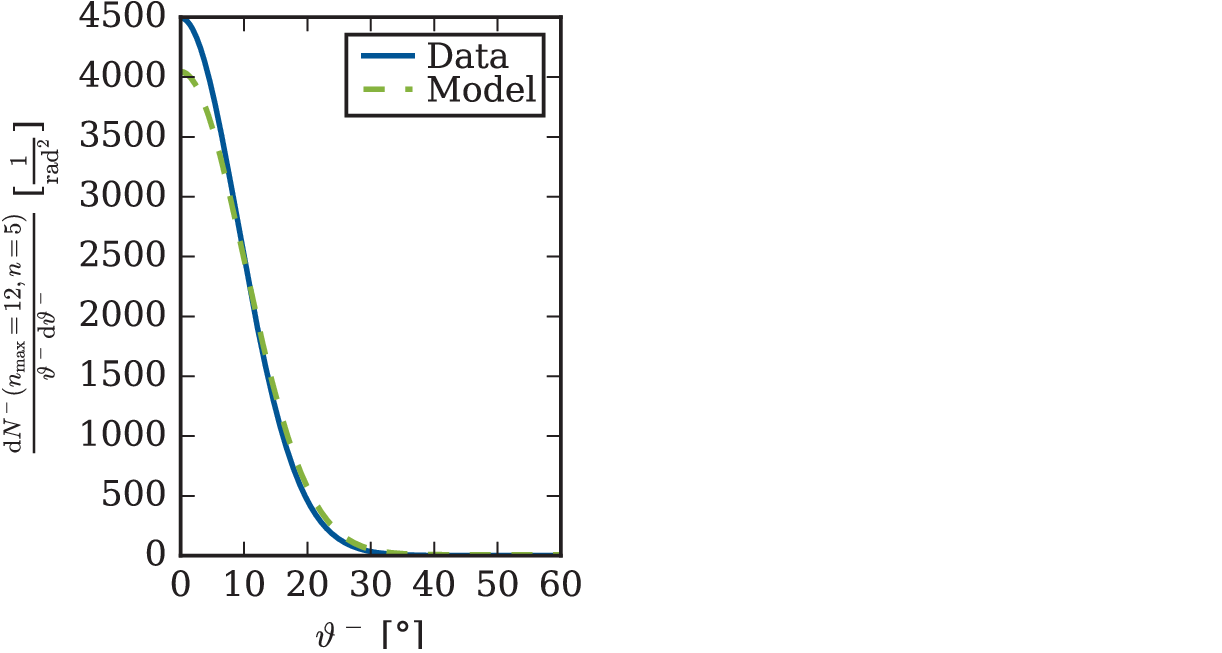}
 \caption{Angular decay of the signal photons emitted in ``$-$'' direction.
 More specifically, here we compare numerical simulation data for the angular decay of $N^-(n_{\rm max}=12,n=5)$ with the respective analytical estimate~\eqref{eq:N-sig}.
 The radial divergence extracted by fitting Gaussian curves to the simulation data, $\theta^-_{\rm sig}(n=5)\simeq\theta$, is in good agreement with the analytical estimate~\eqref{eq:theta-}.}
 \label{fig:angulardecay}
\end{figure}

Figure~\ref{fig:angulardecay} exemplifies the angular decay of the signal photons $N^-$ as inferred from our simulation in comparison to the analytical estimate \eqref{eq:N-sig}.
The corresponding curves are in good agreement.

For the total number of signal photons $N_{(p)}^-(n_{\rm max})=\sum_{n=1}^{n_{\rm max}}N^-_{(p)}(n_{\rm max},n)$ emitted in ``$-$'' direction, \Eqref{eq:N-sig} gives rise to the scaling
\begin{align}
 \frac{N_{(p)}^-(n_{\rm max})}{N_{(p)}^-(n_{\rm max}=1)}
 \simeq  \frac{3}{H_{n_{\rm max}}^{(5/2)}}\sum_{n=1}^{n_{\rm max}}\frac{\sqrt{n}}{2+n^2} \,.
\end{align}
Taking into account Eq.~(11) in the main body of this letter as well as $N_{(p)}^\pm(n_{\rm max}=1)=\frac{1}{2}N_{(p)}(n_{\rm max}=1)$, we obtain 
\begin{align}
\frac{N_{(p)}(n_{\rm max})}{N_{(p)}(n_{\rm max}=1)}
 \simeq \frac{3}{2}\biggl(\frac{1}{n_{\rm max}(1+2n_{\rm max}^2)}\frac{(H^{(1/4)}_{n_{\rm max}})^4}{(H^{(5/2)}_{n_{\rm max}})^2} + \frac{1}{H_{n_{\rm max}}^{(5/2)}}\sum_{n=1}^{n_{\rm max}}\frac{\sqrt{n}}{2+n^2}\biggr) \label{eq:ratioges}
\end{align}
for the total number of attainable signal photons. Equation~\eqref{eq:ratioges} in particular implies $\frac{N_{(p)}(n_{\rm max}=12)}{N_{(p)}(n_{\rm max}=1)}\simeq2.4$.

For completeness, we also provide an analytical estimate for the far-field angular decay of the driving laser photons ${\cal N}^\pm(n)$ in the $n$th mode. Resorting to a paraxial beam model, we obtain \cite{Karbstein:2018omb}
\begin{align}
 \frac{{\rm d}{\cal N}^\pm(n)}{\vartheta^\pm{\rm d}\vartheta^\pm}
 \simeq \frac{4}{\theta^2} \frac{W_n}{n\omega}\,{\rm e}^{-2(\frac{\vartheta^\pm}{\theta})^2} .
\end{align}
For the fundamental-frequency ``$+$'' pulse we have $n=1$.
On the other hand, the ``$-$'' pulse consists of $n_{\rm max}$ modes, i.e., ${\rm d}{\cal N}^-=\sum_{n=1}^{n_{\rm max}}{\rm d}{\cal N}^-(n)$.
The associated laser photon numbers are
\begin{equation}
 {\cal N}^+\simeq \frac{W}{\omega}\quad\text{and}\quad {\cal N}^-\simeq {\cal N}^+\frac{H^{(7/2)}_{n_{\rm max}}}{H^{(5/2)}_{n_{\rm max}}}\leq{\cal N}^+\,,\quad\text{such that}\quad{\cal N}={\cal N}^++{\cal N}^-=\frac{W}{\omega}\biggl(1+\frac{H^{(7/2)}_{n_{\rm max}}}{H^{(5/2)}_{n_{\rm max}}}\biggr)\,.
\end{equation}

\begin{figure}
 \centering
 \includegraphics[width=0.5\linewidth]{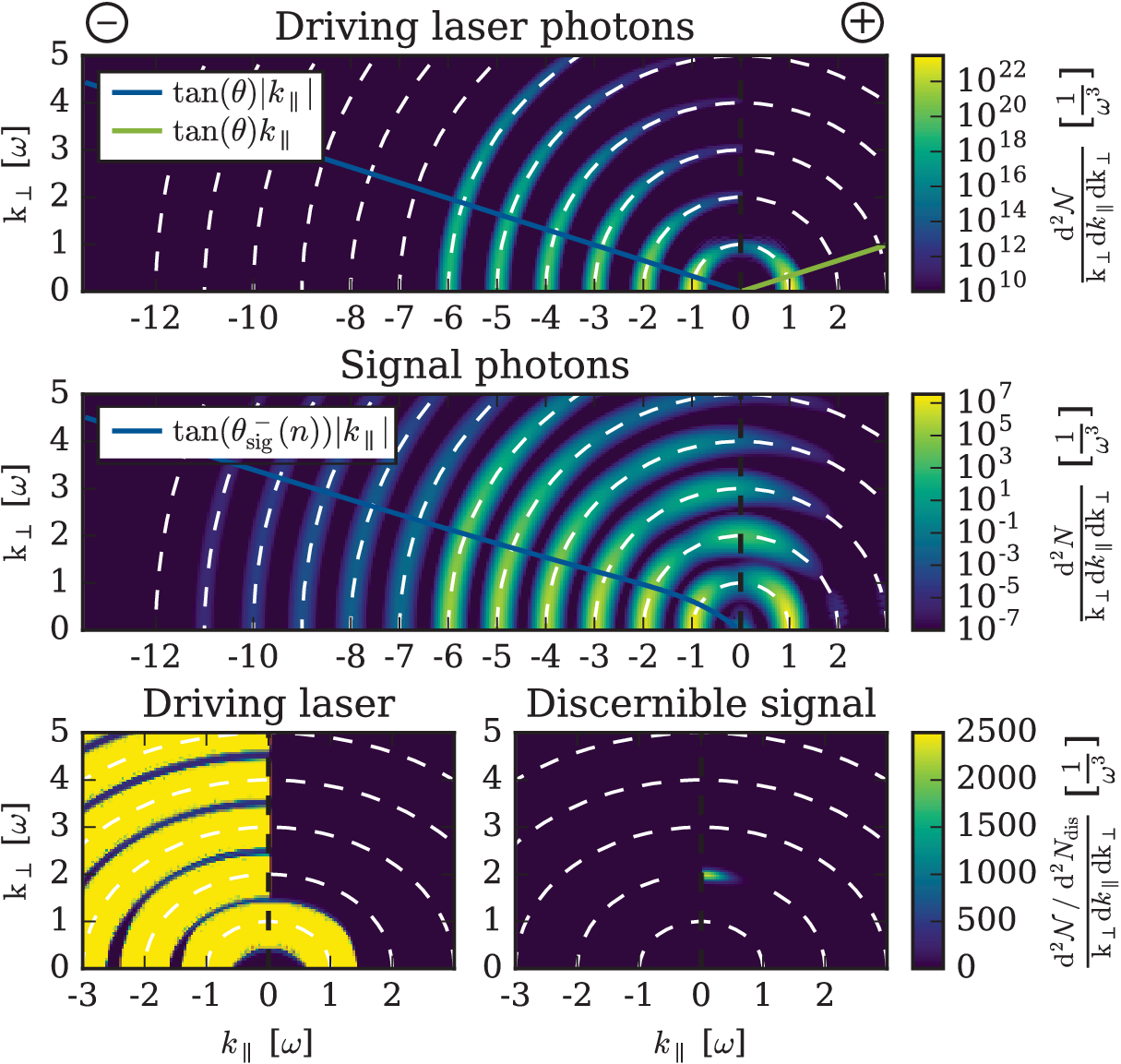}
 \caption{Spectra of the driving laser photons  $\cal N$ and signal photons attainable in a polarization insensitive measurement $N$ for ELI-NP \cite{ELI} laser parameters ($\lambda=800\,{\rm nm}$, $\tau=20\,{\rm fs}$, $W=200\,{\rm J}$) and $n_{\rm max}=6$. The white dashed circles indicate lines of constant photon energy ${\rm k}=n \omega$ with $n\in\mathbb{N}$. Note that different color scales are used in the top, middle and bottom panels.
 For comparison, we also depict analytical estimates for the radial divergences: in the top panel $\theta=1/\pi$ is the radial divergence of a diffraction limited Gaussian beam.
 The radial divergence highlighted in the middle panel is determined from Eq.~\eqref{eq:theta-}.
 The bottom panels focus on the spectral domain where the differential number of signal photons surpasses the differential number of driving laser photons.
 Here, we confront the spectrum of the driving laser photons (left) with the filtered signal photon spectrum fulfilling the criterion ${\rm d}^3N/{\rm d}^3k>{\rm d}^3{\cal N}/{\rm d}^3k$ (right)
 adopting the same linear color scale.
 Integrating the latter, we obtain $N_{\rm dis}\approx314$ \visible\ signal photons per shot at $\approx2\omega$.}
 \label{fig:specsELI}
\end{figure}

Finally, in Fig.~\ref{fig:specsELI} we depict the spectra of the driving laser photons $\cal N$ and signal photons attainable in a polarization insensitive measurement $N$ for the same scenario as discussed in the main text, but assuming ELI-NP \cite{ELI} parameters ($\lambda=800\,{\rm nm}$, $\tau=20\,{\rm fs}$, $W=200\,{\rm J}$) for the driving laser fields and $n_{\rm max}=6$.
This results in $N_{\rm dis}\approx314$ discernible signal photons attainable in a polarization insensitive measurement.
Figure~\ref{fig:specsELI} is qualitatively very similar to Fig.~3 in the main body of this letter.
However, in the figure displayed here the frequency spread of both the various modes constituting the driving laser pulses and the signal photons is substantially smaller.
The reason for this is the larger pulse envelope for the ELI-NP scenario, which is a factor of $4$ larger than the one considered in the main body of this letter.
  

\begin{thebibliography}{10}\setlength{\itemsep}{-0.5mm}

\bibitem{Euler:1935zz} 
  H.~Euler and B.~Kockel,
  Naturwiss.\  {\bf 23}, 246 (1935).

\bibitem{Heisenberg:1935qt} 
  W.~Heisenberg and H.~Euler,
  Z.\ Phys.\  {\bf 98}, 714 (1936), 
  an English translation is available at [physics/0605038].

\bibitem{Weisskopf}
  V.~Weisskopf, 
  Kong.\ Dans.\ Vid.\ Selsk., Mat.-fys.\ Medd.\ {\bf XIV}, 6 (1936).
  
\bibitem{Schwinger:1951nm} 
  J.~S.~Schwinger,
  Phys.\ Rev.\  {\bf 82}, 664 (1951).

\bibitem{Dittrich:2000zu} 
  W.~Dittrich and H.~Gies,
  Springer Tracts Mod.\ Phys.\  {\bf 166}, 1 (2000).
 
\bibitem{Dunne:2004nc} 
  G.~V.~Dunne,
  In *Shifman, M. (ed.) {\it et al.}: From fields to strings, vol. 1* 445-522
  [hep-th/0406216].
 
\bibitem{Marklund:2008gj} 
  M.~Marklund and J.~Lundin,
  Eur.\ Phys.\ J.\ D {\bf 55}, 319 (2009)
  [arXiv:0812.3087 [hep-th]].

\bibitem{Dunne:2008kc} 
  G.~V.~Dunne,
  Eur.\ Phys.\ J.\ D {\bf 55}, 327 (2009)
  [arXiv:0812.3163 [hep-th]].

\bibitem{Heinzl:2008an} 
  T.~Heinzl and A.~Ilderton,
  Eur.\ Phys.\ J.\ D {\bf 55}, 359 (2009)
  [arXiv:0811.1960 [hep-ph]].
  
\bibitem{DiPiazza:2011tq} 
  A.~Di Piazza, C.~M\"uller, K.~Z.~Hatsagortsyan and C.~H.~Keitel,
  Rev.\ Mod.\ Phys.\  {\bf 84}, 1177 (2012)
  [arXiv:1111.3886 [hep-ph]].

\bibitem{Dunne:2012vv} 
  G.~V.~Dunne,
  Int.\ J.\ Mod.\ Phys.\ A {\bf 27}, 1260004 (2012)
  [Int.\ J.\ Mod.\ Phys.\ Conf.\ Ser.\  {\bf 14}, 42 (2012)]
  [arXiv:1202.1557 [hep-th]].
  
\bibitem{Battesti:2012hf} 
  R.~Battesti and C.~Rizzo,
  Rept.\ Prog.\ Phys.\  {\bf 76}, 016401 (2013)
  [arXiv:1211.1933 [physics.optics]].
  
\bibitem{King:2015tba} 
  B.~King and T.~Heinzl,
  High Power Laser Science and Engineering, 4, e5 (2016)
  [arXiv:1510.08456 [hep-ph]].
 
\bibitem{Karbstein:2016hlj} 
  F.~Karbstein,
  arXiv:1611.09883 [hep-th].

\bibitem{Battesti:2018bgc} 
  R.~Battesti {\it et al.},
  Phys.\ Rept.\  {\bf 765-766}, 1 (2018)
  [arXiv:1803.07547 [physics.ins-det]].
  
\bibitem{Lundstrom:2005za} 
  E.~Lundstrom, G.~Brodin, J.~Lundin, M.~Marklund, R.~Bingham, J.~Collier, J.~T.~Mendonca and P.~Norreys,
  Phys.\ Rev.\ Lett.\  {\bf 96}, 083602 (2006)
  [hep-ph/0510076].

\bibitem{Lundin:2006wu} 
  J.~Lundin, M.~Marklund, E.~Lundstrom, G.~Brodin, J.~Collier, R.~Bingham, J.~T.~Mendonca and P.~Norreys,
  Phys.\ Rev.\ A {\bf 74}, 043821 (2006)
  [hep-ph/0606136].

\bibitem{Tommasini:2009nh} 
  D.~Tommasini, A.~Ferrando, H.~Michinel and M.~Seco,
  JHEP {\bf 0911}, 043 (2009)
  [arXiv:0909.4663 [hep-ph]].

\bibitem{Tommasini:2010fb} 
  D.~Tommasini and H.~Michinel,
  Phys.\ Rev.\ A {\bf 82}, 011803 (2010)
  [arXiv:1003.5932 [hep-ph]].

\bibitem{King:2013am} 
  B.~King, A.~Di Piazza and C.~H.~Keitel,
  Nature Photon.\  {\bf 4}, 92 (2010)
  [arXiv:1301.7038 [physics.optics]];
  Phys.\ Rev.\ A {\bf 82}, 032114 (2010)
  [arXiv:1301.7008 [physics.optics]].

  
\bibitem{King:2012aw} 
  B.~King and C.~H.~Keitel,
  New J.\ Phys.\  {\bf 14}, 103002 (2012)
  [arXiv:1202.3339 [hep-ph]].

\bibitem{Gies:2017ygp} 
  H.~Gies, F.~Karbstein and C.~Kohlf\"urst,
  Phys.\ Rev.\ D {\bf 97}, 036022 (2018)
  [arXiv:1712.03232 [hep-ph]].
  
\bibitem{Gies:2017ezf} 
  H.~Gies, F.~Karbstein, C.~Kohlf\"urst and N.~Seegert,
  Phys.\ Rev.\ D {\bf 97}, 076002 (2018)
  [arXiv:1712.06450 [hep-ph]].

\bibitem{King:2018wtn} 
  B.~King, H.~Hu and B.~Shen,
  Phys.\ Rev.\ A {\bf 98}, 023817 (2018)
  [arXiv:1805.03688 [hep-ph]].
  
\bibitem{Blinne:2018nbd} 
  A.~Blinne, H.~Gies, F.~Karbstein, C.~Kohlf\"urst and M.~Zepf,
  Phys.\ Rev.\ D {\bf 99}, 016006 (2019)
  [arXiv:1811.08895 [physics.optics]].

\bibitem{Aboushelbaya:2019ncg} 
  R.~Aboushelbaya {\it et al.},
  arXiv:1902.05928 [physics.optics].

  
\bibitem{CILEX}
  CILEX, \url{http://cilexsaclay.fr/} .
  
\bibitem{CoReLS}
  CoReLS, \url{http://corels.ibs.re.kr/} .

\bibitem{ELI}
  ELI, \url{https://eli-laser.eu/} .

\bibitem{SG-II}
 X.~Xie, J.~Zhu, Q.~Yang, J.~Kang, H.~Zhu, M.~Sun and A. Guo,
 CLEO Technical Digest, paper SM1M.7 (2016).
  
\bibitem{Gordienko:2004} 
  S.~Gordienko, A.~Pukhov, O.~Shorokhov and T.~Baeva,
  Phys.\ Rev.\ Lett.\  {\bf 93}, 115002 (2004).
  
\bibitem{Gordienko:2005zz} 
  S.~Gordienko, A.~Pukhov, O.~Shorokhov and T.~Baeva,
  Phys.\ Rev.\ Lett.\  {\bf 94}, 103903 (2005).

\bibitem{Karbstein:2018omb} 
  F.~Karbstein,
  Phys.\ Rev.\ D {\bf 98}, 056010 (2018)
  [arXiv:1807.03302 [quant-ph]].

\bibitem{Gonoskov:2013ada} 
  A.~Gonoskov, I.~Gonoskov, C.~Harvey, A.~Ilderton, A.~Kim, M.~Marklund, G.~Mourou and A.~M.~Sergeev,
  Phys.\ Rev.\ Lett.\  {\bf 111}, 060404 (2013)
  [arXiv:1302.4653 [hep-ph]].
  
\bibitem{Baeva:2006}
  T.~Baeva, S.~Gordienko and A.~Pukhov,
  Phys.\ Rev.\ E\ {\bf 74}, 046404 (2006).
  
\bibitem{Dinu:2014tsa} 
  V.~Dinu, T.~Heinzl, A.~Ilderton, M.~Marklund and G.~Torgrimsson,
  Phys.\ Rev.\ D {\bf 90}, 045025 (2014)
  [arXiv:1405.7291 [hep-ph]].

\bibitem{Karbstein:2016lby} 
  F.~Karbstein and C.~Sundqvist,
  Phys.\ Rev.\ D {\bf 94}, 013004 (2016)
  [arXiv:1605.09294 [hep-ph]].

  
\bibitem{Karbstein:2014fva} 
  F.~Karbstein and R.~Shaisultanov,
  Phys.\ Rev.\ D {\bf 91}, 113002 (2015)
  [arXiv:1412.6050 [hep-ph]].
  
\bibitem{Gies:2016yaa} 
  H.~Gies and F.~Karbstein,
  JHEP {\bf 1703}, 108 (2017)
  [arXiv:1612.07251 [hep-th]].

\bibitem{Berestetskii}
  V.~B.~Berestetskii, L.~P.~Pitaevskii and E.~M.~Lifshitz,
  Course of Theoretical Physics, Volume 4, Butterworth-Heinemann (Oxford, UK); 2nd edition (1982).
  
\bibitem{Waters:2017tgl} 
  W.~J.~Waters and B.~King,
  Laser Phys.\  {\bf 28}, 015003 (2018)
  [arXiv:1705.08554 [physics.optics]].

\bibitem{Karbstein:2017jgh} 
  F.~Karbstein and E.~A.~Mosman,
  Phys.\ Rev.\ D {\bf 96}, 116004 (2017)
  [arXiv:1711.06151 [hep-ph]].

\bibitem{Laszlo:2017}
  D.~E. Rivas {\it et al.},
  Scientific Reports {\bf 7}, 5224 (2017).

\bibitem{Karbstein:2015xra} 
  F.~Karbstein, H.~Gies, M.~Reuter and M.~Zepf,
  Phys.\ Rev.\ D {\bf 92}, 071301 (2015)
  [arXiv:1507.01084 [hep-ph]].

\bibitem{Karbstein:2019bhp} 
  F.~Karbstein and E.~A.~Mosman,
  Phys.\ Rev.\ D {\bf 100}, 033002 (2019)
  arXiv:1906.10122 [physics.optics].
  
\bibitem{Streeter:2011} 
  M.~J.~V.~Streeter {\it et al.},
  New J.\ Phys.\ {\bf 13}, 023041 (2011).
  
  
\end{thebibliography}
\end{document}